\documentclass[galaxies,article,accept,pdftex,moreauthors]{Definitions/mdpi}

\firstpage{1}
\pubvolume{1}
\issuenum{1}
\articlenumber{0}
\pubyear{2025}
\copyrightyear{2025}
\datereceived{20 May 2025}
\daterevised{19 June 2025} % Comment out if no revised date
\dateaccepted{03 July 2025}
\datepublished{ }
\hreflink{https://doi.org/} % If needed use \linebreak
\usepackage{makecell}

\Title{Quantifying the Unwinding due to Ram Pressure Stripping in Simulated Galaxies}

\TitleCitation{Quantifying the unwinding due to ram pressure stripping in simulated galaxies}

\Author{Rubens E. G. Machado\orcidA{}*, Caroline F. O. Grinberg\orcidB{} and Elvis A. Mello-Terencio\orcidC{}}

\AuthorNames{Rubens E. G. Machado, Caroline F. O. Grinberg and Elvis A. Mello-Terencio}

\AuthorCitation{Machado, R. E. G; Grinberg, C. F. O; Mello-Terencio, E. A.}

\address[1]{Departamento Acad\^emico de F\'isica, Universidade Tecnol\'ogica Federal do Paran\'a, Av. Sete de Setembro 3165, Curitiba, Brazil} 

\corres{Correspondence: rubensmachado@utfpr.edu.br}

\abstract{Galaxies moving through the gas of the intracluster medium (ICM) experience ram pressure stripping, which can leave behind a gas tail. When a disk galaxy receives the wind edge-on, however, the characteristic signature is not a typical jellyfish tail, but rather an unwinding of the spiral arms. We aim to quantify such asymmetries both in the gas and in the stellar component of a simulated galaxy. To this end, we simulate a gas-rich star-forming spiral galaxy moving through a self-consistent ICM gas. The amplitude and location of the asymmetries were measured via Fourier decomposition. We found that the asymmetry is much more evident in the gas component, but it is also measurable in the stars. The amplitude tends to increase with time and the asymmetry radius migrates inwards. We found that, when considering the gas, the spiral arms extend much further and are more unwound than the corresponding stellar arms. Characterizing the unwinding via simulations should help inform the observational criteria used to classify ram pressure stripped galaxies, as opposed to asymmetries induced by other mechanisms.}


\begin{document}

%%%%%%%%%%%%%%%%%%%%%%%%%%%%%%%%%%%%%%%%%%
\section{Introduction}

% environment
Galaxies are affected by their environment. When a gas-rich galaxy travels though the gas of the Intracluster Medium (ICM), it can be stripped of its Interstellar Medium (ISM) gas by the phenomenon of ram pressure \cite{Gunn1972}. This removal of gas may contribute to the quenching of star formation and thus it can have significant effects on the evolution of the galaxy \cite{Boselli2022}.

% ram pressure observations
The morphology of the galaxy is also affected by ram pressure. In extreme cases, a long trail of gas may be left behind. These are know as jellyfish galaxies. To a partial degree, jellyfish-like tails are commonly observed in cluster galaxies. The survey GAs Stripping Phenomena in galaxies (GASP) has been carrying out an intensive program of observations to study ram pressure in galaxies \cite{Poggianti2017, Gullieuszik2017, Jaffe2018}.

% ram pressure simulations
From the simulation point of view, the ram pressure mechanism has been often studied under various conditions, to explore star formation, the effects of inclinations, gas content, orbits etc \cite{Roediger2006, Tonnesen2012, Steinhauser2012, Tonnesen2021, Akerman2023}. Jellyfish galaxies have also been studied within cosmological simulations of galaxy formation, for example, with IllustrisTNG \cite{Yun2019, Joshi2020, Rohr2023} and EAGLE \cite{Troncoso2020, Kulier2023}.

% jellyfish tails versus unwinding
Recently, the peculiar feature of the `unwinding' of the spiral arms has been pointed out in the literature by \citet{Bellhouse2021}. This description of the morphology refers to the shape of the spiral arms, which apparently tend to become more sheared and stretched. This unwinding can be more clearly noticed when the galaxy is seen face-on. Moreover, the configuration which maximizes the unwinding phenomenon is presumably the one where the galaxy receives the wind edge-on. Thus, one side of the disk develops more trailing and open spiral arms, than the side which is moving towards the wind. The layout most favorable to unwinding is the opposite of the more canonical jellyfish configuration. In a classical jellyfish, we seen the disk edge-on, while it receives the wind face-on. In this setting, the galaxy develops a tail ideally perpendicular to the disk itself. The unwinding morphology is therefore less conspicuous than a classical jellyfish. Additionally, it may be confused with tidal tails, which are gravitational in nature; while the origin of the ram pressure induced unwinding is hydrodynamical.

% list previous simulations
In hindsight, such unwinding morphology had indeed manifested itself in a few simulation works \cite{Schulz2001, Roediger2014, Steinhauser2016}. In \citet{Schulz2001}, it was noticed that spirals in the outer disk stretch and shear. In \citet{Roediger2014}, a simulated galaxy with edge-on wind develops strong asymmetry. In \citet{Steinhauser2016}, a degree of asymmetry is also noticeable.

% observations
Observationally, probable unwinding candidates had been seen in \cite{Wolter2015, Bellhouse2017, George2018, Tomicic2018}. Expanding on the topic, \citet{Vulcani2022} performed a visual inspection of $B$-band images and identified 143 galaxies with  unwinding features. More recently, \citet{George2024} presented a prominent example of a strong ram pressure stripped galaxy with unwinding spiral arms, along with additional candidates from GASP. \citet{Krabbe2024} presented methods for finding ram pressure stripped galaxies, and obtained 33 candidates, of which one third showed clear evidence of unwinding arms. \citet{Crossett2025} employed citizen science classifications to identify ram pressure stripped galaxies, and found that those galaxies have a significantly lower than average arm winding with respect to comparison samples.

% open question
Galactic disks may become asymmetric for various other reasons, generally related to tidal interactions. For example, lopsided galaxies are described  as having a distortion where one a side of the disk is more elongated than the other \cite{Lavin2023}; galaxies exhibiting lopsidedness are commonly observed and also studied in simulations. One of the interests in studying the ram pressure induced unwinding is to disentangle the different mechanisms that could lead to similar observed asymmetries. The unwinding of the spiral arms displays morphological features which are reminescent of typical tidal tails. However, there may be characteristic signatures which would discriminate between the different mechanisms. These signatures may be in the kinematics, stellar ages or gas morphology.

% this paper
In this paper, we aim to characterize the properties of a simulated galaxy undergoing ram pressure stripping. In particular, our goal is to quantify the global asymmetry that develops in the galactic disk, and also to explore the phenomenon of the unwinding of the spiral arms. This paper is organized as follows. In section~2 we describe the simulation setup. In section~3 we present the results of our analyses. We conclude with a summary in section~4.

%%%%%%%%%%%%%%%%%%%%%%%%%%%%%%%%%%%%%%%%%%%%%%%%%%
\section{Simulation}

In order to study the phenomenon of ram pressure stripping, we carried out a hydrodynamical $N$-body simulation of a spiral galaxy falling into a cluster. The initial conditions of both the galaxy and the cluster will be described in the following.

% galaxy IC
The galaxy consists of a stellar disk, a gas disk, a bulge and a dark matter halo. The stellar disk follows an exponential profile:
\begin{equation}
\rho_{\rm d}(R,z ) = \frac{M_{\rm d}}{4\pi R_{\rm d}^2 z_{\rm d}} \exp \left( -\frac{R}{R_{\rm d}} \right) {\rm sech}^2 \left( \frac{z}{z_{\rm d}} \right),
\end{equation}
where $M_{\rm d} = 5\times10^{10}\,{\rm M}_{\odot}$ is the mass of the disc, $R_{\rm d} = 3.5$\,kpc is the radial scale length, and $z_{\rm d} = 0.7$\,kpc is the vertical scale length.
For the gas disk, the radial scale length is the same, but the vertical scale length is $z_{\rm g} = 0.035\,z_{\rm d}$. The initial mass of the gas disk is $M_{\rm g} = 2 \times 10^{10}\,{\rm M_{\odot}}$. The bulge and the dark matter halo both follow a Hernquist \cite{Hernquist1990} density profile. The halo mass and scale length are $M_{\rm h} = 1 \times 10^{12}\,{\rm M_{\odot}}$ and $a_{\rm h} = 47$\,kpc. For the bulge, $M_{\rm b} = 1 \times 10^{10}\,{\rm M_{\odot}}$ and $a_{\rm b} = 1.5$\,kpc. These choices of parameters result in a galaxy vaguely similar to the Milky Way.

\begin{figure}
\includegraphics{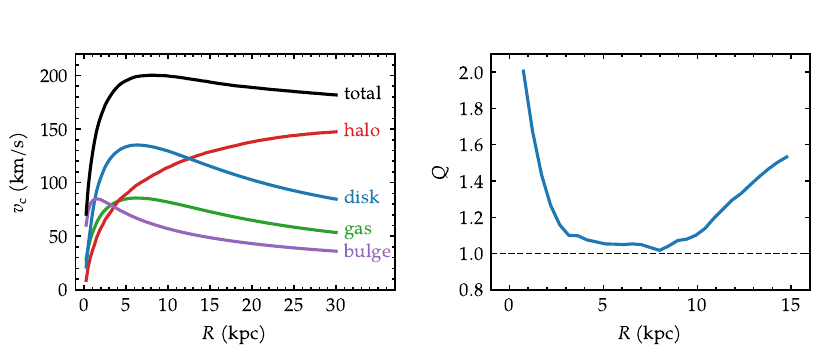}
\caption{Left: Circular velocity curves for the galaxy initial conditions. Right: Toomre parameter of the stellar disk in the initial conditions.}
\label{014}
\end{figure}

% IC kinematics
Fig.~\ref{014} presents the circular velocity curves for the galaxy initial conditions, separated into the components: halo, disk, gas, bulge, and total. The Toomre parameter for the stellar disk is also presented in Fig.~\ref{014}. The method to generate the initial conditions follows the prescription of \citet{SpringelMatteo2005}, where the ratio of radial to vertical velocity dispersions $f_R = \sigma_R^2 / \sigma_z^2$ is a free parameter. We adopted the choice of $f_R=0.8$ as in \citet{Ruggiero2017}, which leads to a relatively stable disc in isolation. In practice, numerical transients occur but dissipate mostly within the first 0.1 Gyr.

% cluster IC
A common approach to simulate ram pressure would be to insert the galaxy into an idealized wind tunnel \cite{Roediger2006, Tonnesen2012}. This approach is more computationally cost-effective but tends to be generally more idealized. Nevertheless, it is possible to set up wind tunnels that emulated the radial density profile of the ICM. Simulating the full ICM self-consistently is another approach that is sometimes adopted \cite{Steinhauser2016, Ruggiero2017}. While it has the advantage of being naturally more realistic, it also much more costly due to the considerable total volume of the ICM. In this work, we opt to create the initial conditions of a self-consistent galaxy cluster. This spherically symmetric galaxy cluster consist of two components: dark matter halo and gas. Both components are represented by the Hernquist \cite{Hernquist1990} profile. The dark matter halo has $M_{\rm h} = 9 \times 10^{13}\,{\rm M_{\odot}}$ and $a_{\rm h} = 280$\,kpc. The ICM gas has $M_{\rm g} = 1 \times 10^{13}\,{\rm M_{\odot}}$ and also $a_{\rm g} = 280$\,kpc. This choice of parameters leads to a galaxy cluster with a dense cool core and a gas fraction of 10\%.

%  codes and number of particles
The initial conditions for the galaxy and for the cluster were created using, respectively, the \textsc{galstep}\footnote{\url{https://github.com/elvismello/galstep}} and the \textsc{clustep}\footnote{\url{https://github.com/elvismello/clustep}} codes \cite{Ruggiero2017}. Realizations were made with the following numbers of particles for the galaxy: $N_{\rm d} = 1 \times 10^5$, $N_{\rm g} = 1 \times 10^5$, $N_{\rm b} = 2 \times 10^4$ and $N_{\rm h} = 2 \times 10^5$, for the disk, gas, bulge and halo, respectively. The cluster was created with $N_{\rm h} = 5 \times 10^6$ particles in the halo and $N_{\rm g} = 5 \times 10^6$ particles in the gas component. These choices imply that the gas component is more finely resolved, which is desirable in the context of the present work. Furthermore, this translates into a mass resolution of $m_{\rm g} = 2 \times 10^5\,{\rm M_{\odot}}$ for the gas particles in the galaxy, and $m_{\rm g} = 2 \times 10^6\,{\rm M_{\odot}}$ for the gas particles in the cluster. In this multiscale setup, we are faced with a wide dynamical range between the mass of the galaxy and the mass of the cluster. Thus, the computational cost of a fixed mass resolution across all particles would have been exceedingly challenging in practice.

% resolution caveats
The mass resolution of this simulation may be contrasted to those in IllustrisTNG simulations, a well established state-of-the-art simulation suite in the field. For instance, TNG100 adopts a baryonic mass resolution of $1.4 \times 10^6\,{\rm M}_\odot$ \cite{Springel2018}, and has been successfully used to study jellyfish galaxies, including detailed analyses of ram pressure interactions between the interstellar medium and intracluster medium, as well as the resulting morphological signatures \citep{Yun2019}. Yet, important caveats remain. The sensitivity of galactic dynamics to numerical resolution is a common concern. For example, \citet{Sellwood2012, Sellwood2013} showed how particle resolution can significantly affect secular processes such as bar formation and disk heating. Moreover, even in high-resolution cosmological simulations such as TNG50 \cite{Nelson2019}, some dynamical phenomena remain insufficiently captured. One example is the so-called fast bar tension discussed in \cite{Habibi2024}, which persists in TNG50 but appears to be alleviated in the higher-resolution Auriga simulations \citep{Fragkoudi2021}. Thus, while our adopted resolution is consistent with values commonly used in the field, this does not eliminate the possibility of numerical effects on small-scale dynamics. A systematic convergence study is deferred to a future work.

% join
At the start of the simulation ($t=0$), the galaxy is placed at $x=1000$\,kpc, with a velocity of $v_x = -1000$\,km\,s$^{-1}$ entirely along the $x$ axis. In other words, the galaxy is set to fall in a radial orbit towards the center of the cluster. Since the galactic disk lies on the $xy$ plane, and the velocity vector of the galaxy is confined to the same plane, this means that the galaxy will receive the wind always edge-on throughout the simulation.

% code 
The simulation is performed with Gadget-3, a version of the Gadget-2 code \cite{Springel2005} including star formation. The maximum integration time step was 0.001\,Gyr. Before the simulation starts, the dark matter particles of the cluster are excised from the snapshot. Instead, the gravitational potential of the cluster halo is represented by a frozen analytic potential with the exact same parameters as the $N$-body realization. This is justified by the fact that, during the evolution, the dark matter of the cluster will not be significantly altered by the passage of the galaxy. Replacing the $N$-body particles by an analytic potential reduces computational cost.

\begin{figure}
\includegraphics{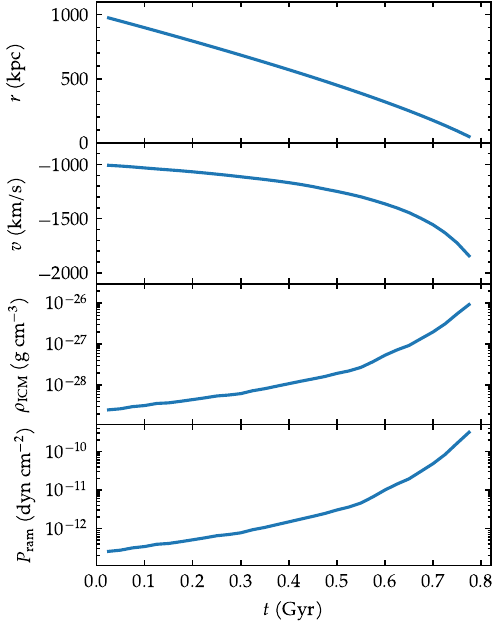}
\caption{Orbital evolution of the galaxy. The first and second panels display the radial position and the velocity of the galaxy. The third and fourth panels show the density of the ICM gas it encounters, and the corresponding ram pressure.}
\label{fig1}
\end{figure}

\begin{figure}
\includegraphics{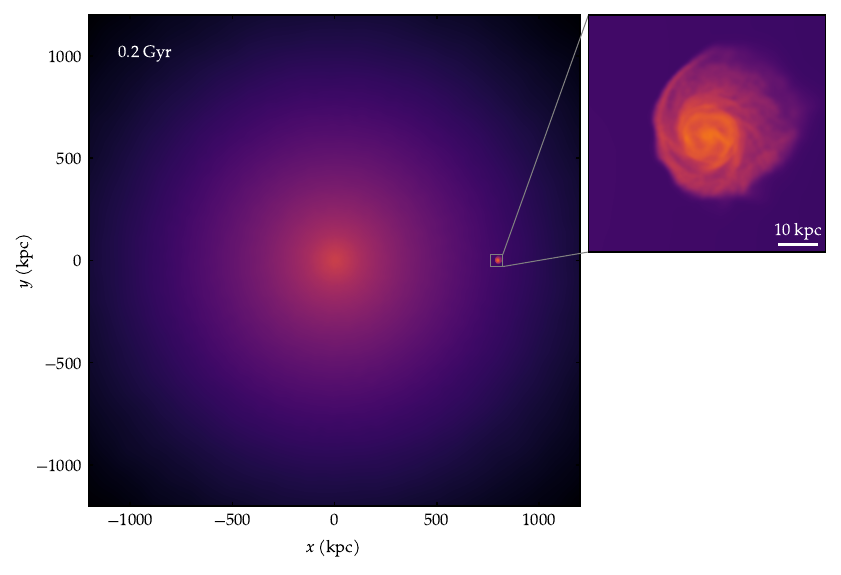}
\caption{Map of projected gas density showing the galaxy falling into the cluster, at $t=0.2$\,Gyr.}
\label{fig2}
\end{figure}

% fig1
The simulation is carried out for 1\,Gyr, which is roughly the time required for the galaxy to traverse the radius of the cluster. More precisely, the first two panels of Fig.~\ref{fig1} indicate the evolution of the orbital parameters of the galaxy, namely its clustercentric distance and its radial velocity as a function of time. Throughout the paper, the time evolution will be generally presented up to $t\sim0.8$\,Gyr at most, or shortly earlier. At later times, the galaxy would cross the very center of the potential well, leading to unphysical consequences. Therefore, our discussions are limited to the infall from $r=1000$\,kpc until the galaxy reaches about $r\sim$100\,kpc.

% Pram equation
As the galaxy travels towards the center, it meets progressively denser gas. Fig.~\ref{fig1} also displays the density of the ICM gas which is encountered by the galaxy at each time in its trajectory, as well as the ram pressure that acts on the galaxy. From the gas density $\rho_{\rm ICM}$ and the wind velocity $v$, the ram pressure is computed as:
\begin{equation}
P_{\rm ram} = \rho_{\rm ICM} ~v^2.
\end{equation}

% fig2
An illustrative visualization of the galaxy in the presence of the cluster environment is given in Fig.~\ref{fig2}, where the colors represent the gas density of both components together, ISM and ICM. This example is shown at $t=0.2$\,Gyr, by which time the galaxy has already fallen a few 100\,kpc. The zoomed-in inset panel in Fig.~\ref{fig2} helps underscore the disparity of spatial scales involved in this problem.

%%%%%%%%%%%%%%%%%%%%%%%%%%%%%%%%%%%%%%%%%%%%%%%%%%
\section{Analysis and results}

\subsection{Global evolution}

\begin{figure}[H]
\includegraphics[scale=0.96]{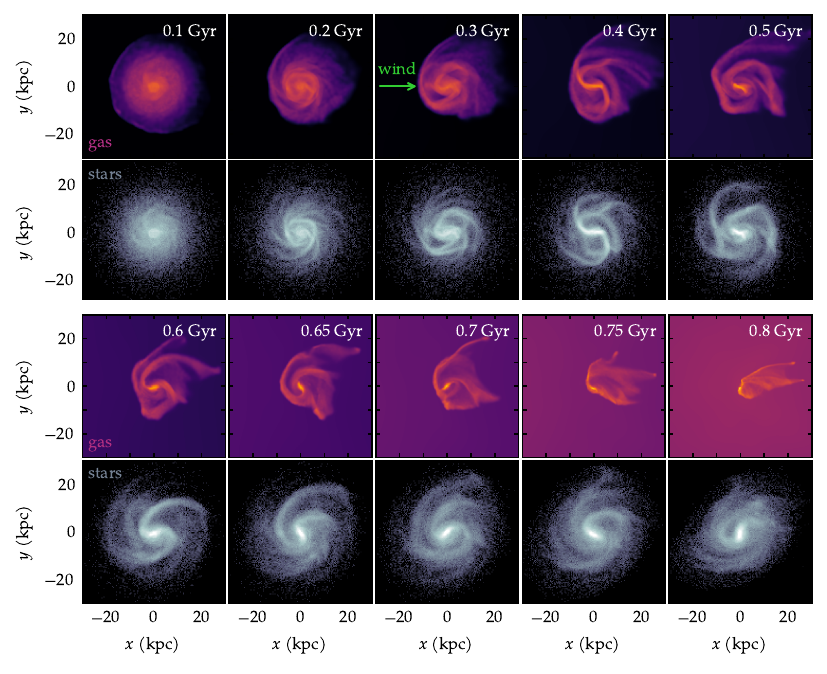}
\caption{Maps of projected density of gas and stars at selected times during the evolution. Notice that the background ICM gas is also shown. The green arrow represents the constant wind direction.}
\label{fig3}
\end{figure}

% overall evolution
The effects of ram pressure become noticeable as soon as the simulation starts --- mildly at first, and then strongly as time progresses. Fig.~\ref{fig3} exhibits the morphological evolution of the gas and of the stars, at selected times. In this and in all other subsequent visualizations of the paper, the origin of the coordinate system has been shifted to coincide with the center of the galaxy. In the reference frame of the galaxy, the wind direction is to be understood as arriving from the left towards the right. This wind direction is constant throughout the simulation and is shown as a green arrow in one of the panels of Fig.~\ref{fig3}.

% early phase
Already at $t=0.1$\,Gyr, the leftmost side of the gas disk is undergoing a subtle but noticeable amount of compression. At the same time, the stellar disk remains fairly axisymmetric. Between 0.1 and 0.5\,Gyr, the ram pressure grows steadily, and as a results the gas disk becomes clearly asymmetric. This is seen in the upper set of frames in Fig.~\ref{fig3}. The gradual unwinding of the spiral arms is qualitatively perceptible from this visualization.

\begin{figure}[H]
\includegraphics[scale=0.96]{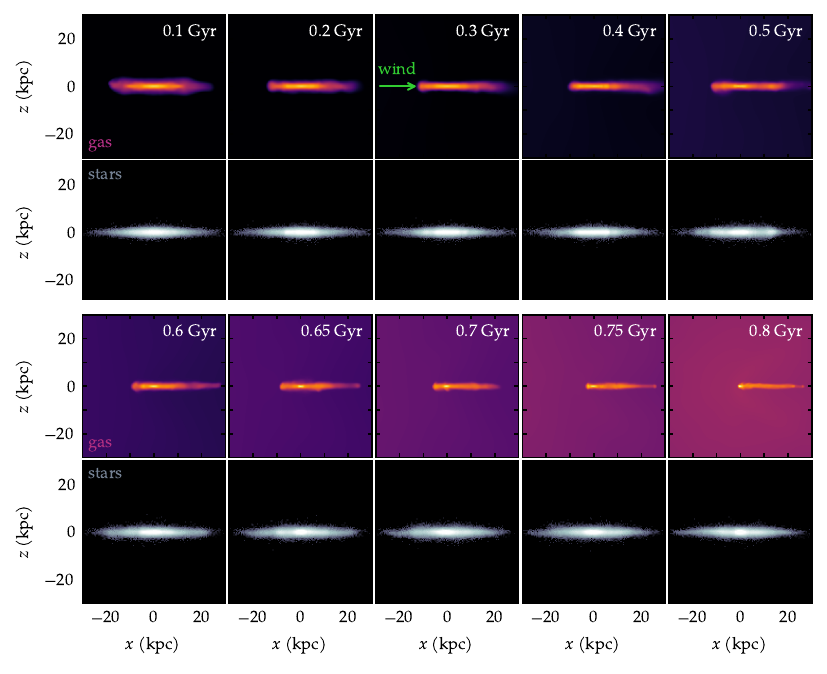}
\caption{Same as Fig.~\ref{fig3} but showing an edge-on view.}
\label{017}
\end{figure}

% later pahse
In the lower set of frames in Fig.~\ref{fig3}, selected times are shown in the range from 0.6 to 0.8\,Gyr. This later phase corresponds to a distinct regime. Firstly, the time scale itself is naturally shorter, due to the acceleration of the galaxy. The evolution of the velocity seen in Fig.~\ref{fig1} indicates that this is the phase of greatest acceleration, as the galaxy approaches the core of the cluster. Likewise, both the density and the ram pressure are intensified at this final phase. According to Fig.~\ref{fig1}, after about $t\sim0.55$\,Gyr, there is a marked change in the slope of the evolution of $\rho_{\rm ICM}$ and of $P_{\rm ram}$. By this time, the galaxy has already travelled some $\sim$600\,kpc. Yet, during the remaining $\sim$300\,kpc, it will experience an increase of ram pressure more pronounced than in the earlier phase. The consequence of this extreme ram pressure is seen in the 0.6--0.8\,Gyr gas panels of Fig.~\ref{fig3}. Inspection of slightly later snapshots (not shown) confirms that the gas tail seen at 0.8\,Gyr will be mostly stripped away. After 0.8\,Gyr, both the gas and the stars undergo distortions that would not be meaningful to interpret physically. In a real cluster in the Universe, a collision would probably have occurred with a Brightest Cluster Galaxy (BCG) by that point. 

% very late phase
It should also be noted that the apparent asymmetry of the stellar disk towards $t\sim0.8$\,Gyr cannot be confidently attributed to ram pressure effects, and is more likely due to strong tidal forces near the cluster center. In this sense, interpretations of both the gas and the stars should be regarded with caution at the latest snapshots. For these reasons, most of the analysis will focus on the more reliable early phases. The largely unwound gas spiral arms themselves are a disturbance that could in principle have some local influence on the nearby stars. Even though the gas mass is relatively small, it may exert subtle tidal forces on the stellar arms.

Fig.~\ref{017} displays the equivalent evolution of Fig.~\ref{fig3}, but showing an edge-on view. Since the galaxy receives the wind in a direction entirely parallel to the disk, there is no cause for vertical asymmetries. As a result, neither the gas disk nor the stellar disk develop any noticeable features in the vertical direction. The unwound spiral arms in the gas are essentially confined to the same heights as the rest of the gas disk.

\subsection{Amplitude and location of the asymmetries}

% Fourier
We wish to quatify the degree to which the morphology of the galaxy departs from an axisymmetric configuration. Strictly speaking, once spiral arms are well developed, the disk is no longer axisymmetric, even in an unperturbed galaxy. Nevertheless, the phenomenon of unwinding clearly produces a global elongation in a preferred direction. In other words, beyond a given radius, there is more material (more extended arms) in one side of the galaxy than in the opposite side. This feature is suitable to be quantified by means of a Fourier decomposition of the projected mass distribuition. In particular, the $m=1$ mode should be able to capture the signal of such an asymmetry. The relevant coefficients can be written as:
\begin{eqnarray}
a_0 &=& \sum_{i=0}^N ~ m_i \\
a_1 &=& \sum_{i=0}^N ~ m_i ~ \cos\theta_{i} \\
b_1 &=& \sum_{i=0}^N ~ m_i ~ \sin\theta_{i}\,,
\end{eqnarray}
where $m_i$ is the mass of each particle and $\theta_i$ is the azimuthal angle of each particle. The summation is performed using all the $N$ particles that fall inside a given annulus. Thus, we can define the amplitude $A_1$ as:
\begin{equation}
A_1 = \frac{\sqrt{a_1^2+b_1^2}}{a_0}
\end{equation}
and measure it as function of cylindrical radius $R$.

\begin{figure}
\includegraphics{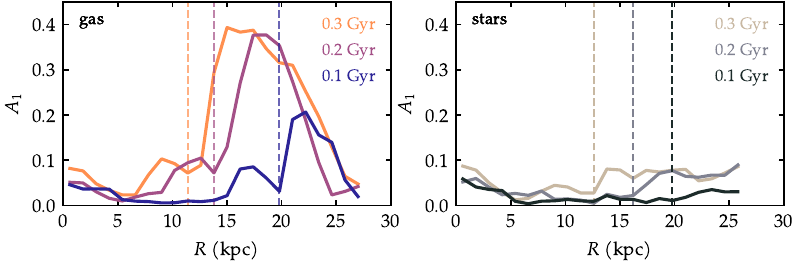}
\caption{Radial profiles of $A_1$ for the gas and the stars at different times. The vertical lines correspond to the radii of the minimum value of $A_1$ before the peak; this is the asymmetry radius $R_{\rm s}$.}
\label{fig4}
\end{figure}

% A1 as a function of radius, gas
Fig.~\ref{fig4} presents the results of this measurement performed for both the gas and the stars at different times. The profiles are generally noisy because this metric also captures the presence of the spiral arms themselves. Nevertheless, when a very important asymmetry is present, it is clearly seen as a peak in the radial profile of $A_1(R)$. In the left panel of Fig.~\ref{fig4}, we see that the peaks in the gas are quite pronounced. As time progresses, the peaks seem to become higher and also move towards smaller radii. (To ensure that the asymmetry is not due to mere secular evolution, the measurement was repeated for an isolated galaxy; see Appendix~\ref{app:isolated}.) The location of the peaks has a direct interpretation: it is the radius of greatest asymmetry. These peaks are also useful to define a certain radius: the radius within which the galaxy is still relatively `symmetric', or put another way, the radius beyond which the asymmetry becomes very pronounced --- we define it as the radius $R_{\rm s}$. This radius could be referred to as the `asymmetry radius' in the sense that that is where the asymmetry begins. This radius $R_{\rm s}$ is measured in practice as the first minimum to the left of the peak of $A_1(R)$. The vertical lines in Fig.~\ref{fig4} mark the values of $R_{\rm s}$ obtained for the corresponding times shown in the figure. 

\begin{figure}
\includegraphics{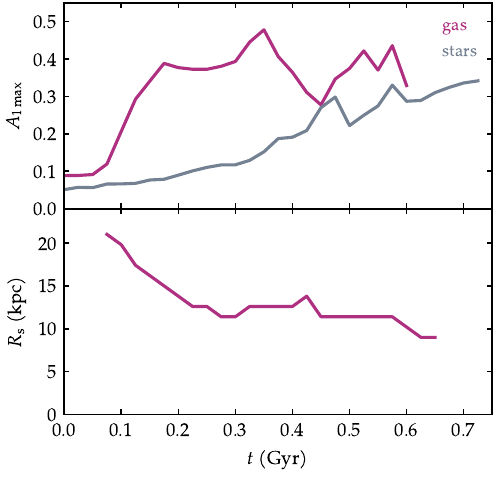}
\caption{Top: time evolution of $A_{\rm 1 max}$ for the gas and the stars. Bottom: time evolution of the asymmetry radius, as measured from the gas. }
\label{fig5}
\end{figure}

% A1 as a function of radius, stars
The same procedure that was done for the gas can also be applied to the stars. This is shown in the right panel of Fig.~\ref{fig4}. The radial profiles show that the amplitude of $A_1$ for the stars is much smaller, but non-zero. It is also true that the amplitude increases with time. The $R_{\rm s}$ are in principle measurable, but are ill-defined and difficult to determine.

% A1max
At each time in the simulation, we can measure the height of the peak in the radial profile of $A_1$. This is the definition of the quantity $A_{1 \rm max}$, which is shown in the upper panel of Fig.~\ref{fig5}, for gas and stars. Here we see a more quantitative characterization of the asymmetries and their evolutions. The asymmetry of the stars is smaller and evolves slowly. The asymmetry of the gas is more pronounced and grows quickly in the beginning of the simulation. At later times, the comparison becomes less conclusive. At times when the morphology of the spiral arms is extremely irregular and peculiar, $A_1$ captures these signatures rather than the intended global asymmetry.

% RS
The evolution of the asymmetry radius $R_{\rm s}$ is shown in the lower panel of Fig.~\ref{fig5} only from the gas measurements, because in the stellar component the profiles of $A_1(R)$ are too noisy to find $R_{\rm s}$ reliably. It is interesting to notice that $R_{\rm s}$ decreases with time. This means that, at first, the asymmetry is manifested only in the outskirts of the disk, while most of the galaxy remains relatively symmetric. As time progresses, the asymmetry radius migrates inward, meaning that now a larger part of the outer disk is disturbed as well.

\subsection{Direction of the elongation}

\begin{figure}
\includegraphics{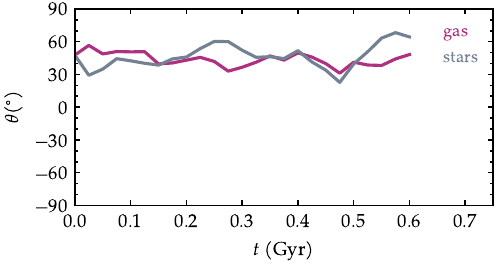}
\caption{Direction of the elongation, as determined by the Radon transform, for both gas and stars.}
\label{fig6}
\end{figure}

% elongation
The galactic disk undergoing ram pressure under these conditions is no longer circular, but rather elongated. The elongation is somewhat analogous to the `tail' of a jellyfish. However, in a more classical jellyfish, the tail is generally expected to point in the direction opposite to the direction of motion of the galaxy; it is a wake of material left behind. In the edge-on case analyzed here, a striking feature seen in the morphology of all snapshots is that the major axis of this elongation is not parallel to the wind direction. Instead, there is an angle between the direction of the wind and the direction of the elongation. This angle seems to be roughly constant and seems to be connected to the sense of rotation of the galaxy.

% Radon
In order to quantify the direction of the elongation, we chose to apply a Radon transform, a tool occasionally used in Astronomy \cite{Stark2018}; see footnotes in \cite{Krone2013}. The Radon transform of an image results in a so-called `sinogram' whose most intense projection gives the preferred direction. The definition is given in Appendix~\ref{app:radon}. This was measured as a function of time for both gas maps and stellar maps. The results are shown in Fig.~\ref{fig6}. The general conclusion is that the elongation (i.e.~the `tail' equivalent) of the galaxy points roughly $\sim45^{\circ}$ away from the motion of the galaxy. The measurement of this angle is likewise hindered by the irregularity of the spiral arms themselves, resulting in a relatively broad dispersion of some $\pm10^{\circ}$ around the average over time. Nevertheless, it is remarkable that the elongation is consistently confined to (the middle of) the first quadrant. In all visualization images in this paper, the rotation of the galaxy is counter-clockwise. It seems consistent that the leading edges of the spiral arms become preferentially compressed when they meet the wind in the second quadrant.

\begin{figure}[ht]
\includegraphics{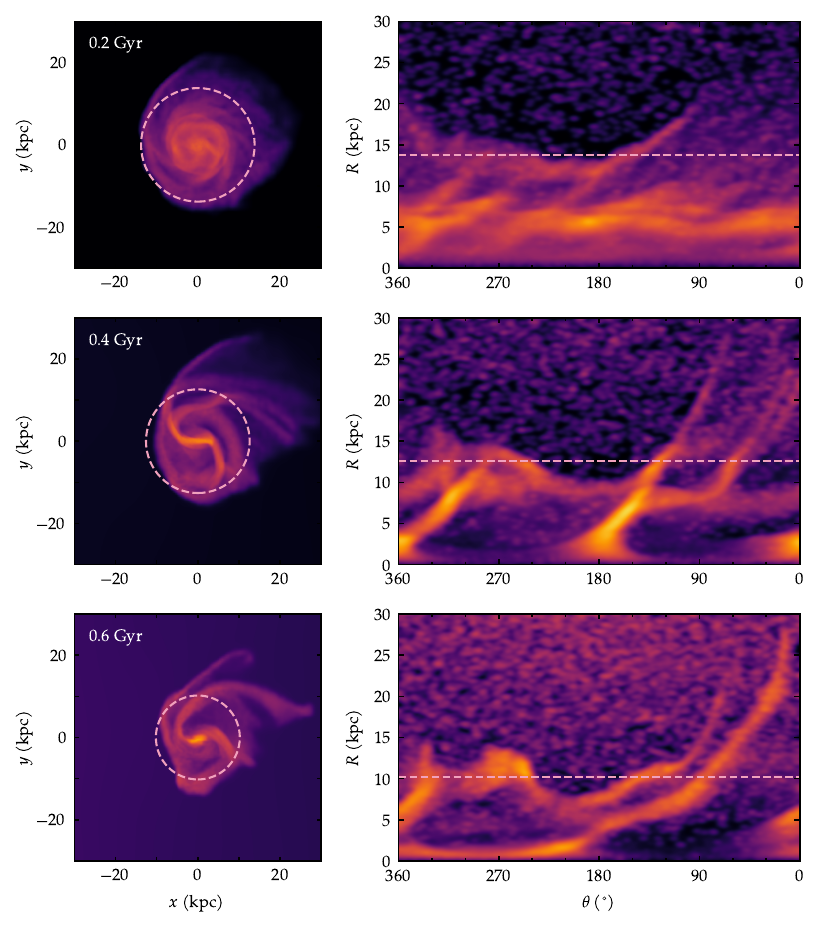}
\caption{Left: maps of projected gas density. Right: azimuthal plots at the corresponding times. The circles and the horizontal lines correspond to the asymmetry radius $R_{\rm s}$.}
\label{fig7}
\end{figure}

\subsection{Shapes of the spiral arms}

\begin{figure}[t]
\includegraphics{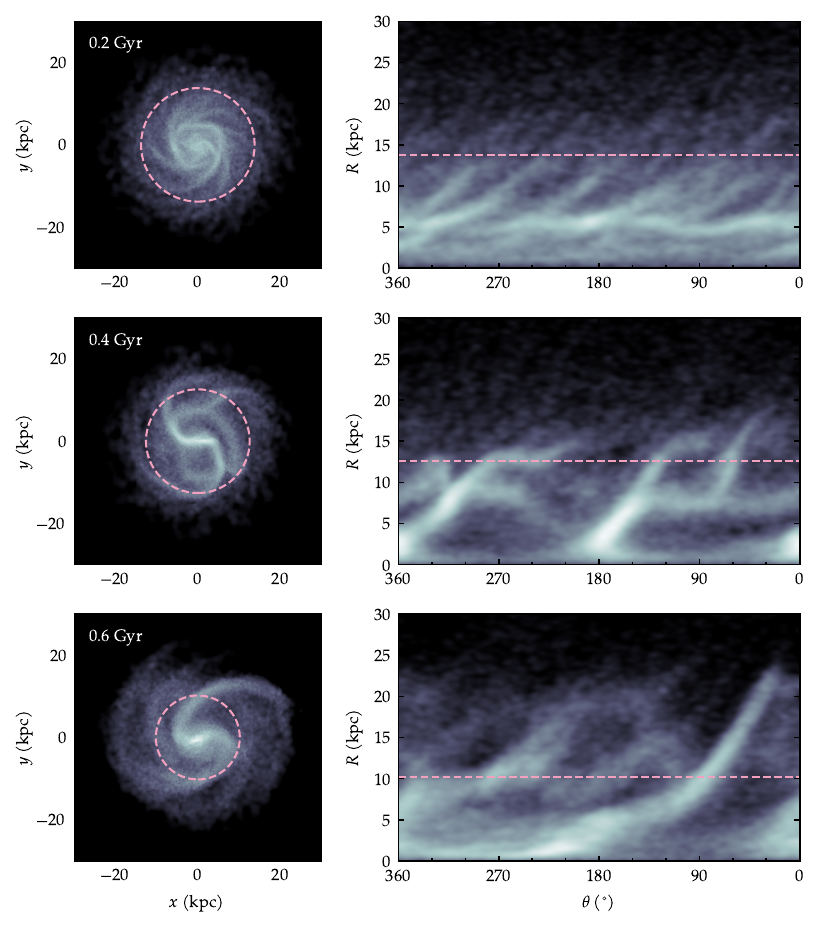}
\caption{Left: maps of projected stellar density. Right: azimuthal plots at the corresponding times. The circles and the horizontal lines correspond to the asymmetry radius $R_{\rm s}$.}
\label{fig8}
\end{figure}

% RS circles gas
Finally, in order to scrutinize the shapes of the spiral arms more closely, we unwrap the azimuthal coordinate in the form of rectangular plots of the polar coordinates: $R$ as a function of $\theta$. This is shown separately for the gas in Fig.~\ref{fig7}, and for the stars in Fig.~\ref{fig8}. Accompanying the azimuthal plots are normal density maps of the galaxy at the corresponding times. The dashed pink circles overlaid on the density maps correspond to the asymmetry radius $R_{\rm s}$. It is noticeable in the gas maps of Fig.~\ref{fig7} that the circles indeed enclose the region which could be considered relatively symmetric, thus highlighting the outer region which clearly lacks symmetry.

% RS circles stars
In the stellar density maps of Fig.~\ref{fig8}, the same pink circles are shown, i.e.~the same $R_{\rm s}$ measured from the gas, not the stars. As had been pointed out earlier, even though the stellar asymmetry can be quantitatively detected by $A_1$, it is not qualitatively visible by eye.

% explain azimuth plot
Turning to the azimuthal plots on the right panels of Fig.~\ref{fig7}, the corresponding values of $R_{\rm s}$ naturally translate into a line of constant radius in the $R-\theta$ plot. This helps guide the view of the outermost part of the galaxy, where the asymmetries are most important. The horizontal axes of the azimuthal plots are shown counter-clockwise (from 360$^{\circ}$ to 0$^{\circ}$), to help guide the eye as one follows a given spiral arm along its azimuths. One can read these plots in the following manner. If a given spiral arm were very tightly wound --- or in the limiting case, if it were a ring --- then, as one reads its azimuths, it would trace a region of roughly constant $R$, i.e.~a flat feature in the $R-\theta$ plot. On the other hand, a very open unwound arm reaches increasingly larger values of $R$ as one sweeps its azimuths; thus an unwinding arm translates into an upward branch in the $R-\theta$ plot.

% describe fig7 branches
With this interpretation in mind, we can appreciate in Fig.~\ref{fig7} the progressive unwinding of the arms. At first, at $t=0.2$\,Gyr, there is hardly any upward branch in the azimuthal plot, apart from a short segment peaking above the dashed line. At $t=0.4$\,Gyr, there are tenuous but very long branches reaching up above the dashed line. At $t=0.6$\,Gyr, there are stronger branches. Notice that the preferential region for the upward branches is the first quadrant (i.e. between 90$^{\circ}$ and 0$^{\circ}$). This is consistent with the general elongation of the galaxy being confined to this quadrant, as seen in the previous subsection.

% describe fig8 branches
Now, contrasting the gas to the stars of Fig.~\ref{fig8}, we find interesting differences. At $t=0.2$\,Gyr, there are no unwinding branches, just common spiral arms covering azimuths uniformly. At $t=0.4$\,Gyr, we see that there is a correspondence between the main arms in gas and in stars. However, the faint thin tendrils seen in the gas are missing in the stars. At $t=0.6$\,Gyr, one arm is present in the gas but absent in the stars; another arm extends further in the gas, while it is truncated sooner in the stars.

% density maps figs 7,8
The nature of these features is confirmed when comparing directly the density maps Figs.~\ref{fig7} and ~\ref{fig8}. We find that within the circle of radius $R_{\rm s}$, there is a general match between spiral arms seen in the gas and in the stars. However, beyond $R_{\rm s}$ there are clear departures in the gas, whose arms extends much further and are more unwound than the corresponding stellar arms.

%%%%%%%%%%%%%%%%%%%%%%%%%%%%%%%%%%%%%%%%%%%%%%%%%%
\section{Discussion}

% results: global
In this paper, we have analyzed a hydrodynamical simulation in which a galaxy falls into a cluster, encountering the wind edge-on. In this configuration, the ram pressure causes the spiral arms to unwind. During the first phase of the infall, the ram pressure builds up gradually, stripping the gas and causing the gas to become noticeably asymmetric. At a later phase, as the galaxy approaches the denser gas in the core of the cluster, the ram pressure steepens significantly and the spiral arms in the gas are swept away almost completely. Meanwhile, the effects on the stellar disk are more subtle.

% results: amplitude of asymmetries
We have quantified the amplitude and the location of the asymmetries by a Fourier decomposition of the projected mass. We found that the asymmetry is much stronger in the gas, but it is also detectable in the stellar disk as well. The amplitude of the asymmetry generally increases over time. In the case of the stars, this evolution is slow and mild. In the case of the gas, the asymmetry grows steeply in the beginning.

% results: amplitude of asymmetries
Furthermore, we quantified the radius beyond which the asymmetries are most important. In general, the phenomenon of ram pressure tends to impact first the gas in the periphery of the disk, which is more precariously bound to the galactic potential. That is also the case here. The asymmetry radius is large at first, meaning that only the outskirts beyond 20\,kpc are affected in the first phase. At later times, this radius migrates inwards to 10\,kpc, showing that a relevant fraction of the disk is perturbed.

% results: direction
Even though the galaxy was set to move along the $x$ axis, we found that the direction of the elongation (analogous to a `tail') was on average at an angle of $\sim45^{\circ}$ with respect to the line of motion. This is consistent with the findings of \citet{Bellhouse2021}. This tail direction depends of the motion of the galaxy and also on its sense of rotation, as on one side the leading part or the arms moves against the wind, and on the the other side, with the wind. This results is potentially interesting to clarify the probable direction of the trail vectors in certain jellyfish galaxies. In the case of galaxies the receive the wind face-on, the direction of the tail is generally interpreted as pointing in the direction opposite to the velocity vector of the galaxy. In the case of a ram pressure stripped galaxy moving edge-on, the direction of motion is less clear.

% results: azimuths
Following the approach of \citet{Bellhouse2021}, we employed azimuthal diagrams in order to characterize the unwinding of the spiral arms. This is a useful tool to inspect the morphological evolution of the spiral arms. We found that the upward branches in the diagrams, representing the unwinding arms, are the longest and the most intense in the first quadrant. Additionally, the features in the gas are seen to extend further than in the stars.

% previous work
In comparison to previous works, the type morphology that arises in our simulations is qualitatively similar. For example, \citet{Schulz2001} had seen arms that were stretched and sheared. \citet{Roediger2006, Roediger2007} had simulated galaxies undergoing ram pressure under different inclinations and found a strong degree of asymmetry in the $xy$ plane in an equivalent wind configuration. Such noticeable edge-on stripping was also presente in \citet{Roediger2014}. It is interesting to notice that the conditions under which such asymmetry develps are apparently broad. More quantitatively, the analysis of \citet{Bellhouse2021} further explored the detailed morphology of the spiral arms, both in observations and in simulations. Our simulation confirms that a galaxy falling into a self-consistent cluster will develop a similar degree of unwinding in the gas, as seen from the azimuthal diagrams.

% why stars are affected
A relevant issue to address is the question of why the stellar component should be affected at all, since ram pressure is in principle a purely hydrodynamical effect. In this regard, the main effect expected to be at play is star formation. Since the gas is continually forming stars in the simulation, star formation takes place also in the recently disturbed gas spiral arms. For this reason, the stellar spiral arms should be at least partially perturbed, alhtough the bulk of the older stellar population would remain mostly undisturbed. This expectation is in line with the results of \citet{Bellhouse2021}, who find predominantely young stars in the unwound component. A secondary effect might also contribute to the stellar asymmetries, to a lesser degree. When the gas disk is strongly stripped by ram pressure, this redistribution of mass may affect the stellar component through their mutual gravity. This has been suggested by ram pressure simulations \cite{Smith2012} in which the stellar disk is temporarily deformed, and even the central region of the dark matter halo becomes slightly offset. Such indirect efects might also play a role in perturbing the stellar spiral arms in the edge-on wind case.

% caveats
The environment through which the galaxy travels determines the impact of the ram pressure stripping. This is why the specific orbits are important \cite{Tonnesen2019}. In this paper, only one galaxy trajectory was considered, namely a radial orbit; and only one inclination, edge-on. These choices were made in order to obtain a configuration in which the effect of interest would be most pronounced. Moreover, the density profile of the ICM gas exhibits a dense cool core. By design, these environmental and orbital properties all contribute towards our simulated galaxy having undergone a particularly intense history of ram pressure stripping.

\section{Conclusions}

Our conclusions can be summarized as follows. (i) A galaxy receiving the wind edge-on undergoes unwinding of the spiral arms. In agreement with previous works, we find that this phenomenon occurs naturally in this circumstance. Additionally, we confirm that the direction of the tail tends to be on average at an angle of $\sim45^{\circ}$. (ii) This works extends the analysis to the stellar component as well. We find that the unwinding of the spiral arms is also measurable in the stars, although it is much more noticeable in the gas. In the case of the gas, arms extend further out and are more unwound.

% perspectives
In future simulations, it would be interesting to explore the unwinding phenomenon in galaxies moving along different environments and receiving the wind under varied inclinations. In particular, the present analysis could be repeated for populations of different stellar ages, in order to confirm the expected predominance of young stars in the unwound component. From a theoretical standpoint, these kinds of simulations can be used to explore in greater detail the dynamics of the gas to understand how the velocities in the spiral arms are impacted by the compression of the gas. Furthermore, this paper was focused on the density of the gas and stellar components. In order to offer a closer connection to observations, a useful perspective would be to produce mock images in specific observational bands and in H$\upalpha$. These would help with the interpretation of the observations of ram pressure stripped galaxies exhibiting unwinding of the spiral arms.

%%%%%%%%%%%%%%%%%%%%%%%%%%%%%%%%%%%%%%%%%%
\vspace{2pt}

\authorcontributions{Conceptualization, RM; methodology, RM, CG, EMT; software, RM, CG, EMT; formal analysis, RM, CG; resources, RM; writing---original draft preparation, RM; writing---review and editing, RM; visualization, RM; supervision, RM; project administration, RM; funding acquisition, RM. All authors have read and agreed to the published version of the manuscript.}

\funding{RM acknowledges support from the Brazilian agency \textit{Conselho Nacional de Desenvolvimento Cient\'ifico e Tecnol\'ogico} (CNPq) through grants  406908/2018-4, 303426/2018-7, and 307205/2021-5 and from \textit{Fundação Araucária} through grant PDI 346/2024 -- NAPI \textit{Fenômenos Extremos do Universo}. CG acknowledges support from \textit{Coordenação de Aperfeiçoamento de Pessoal de Nível Superior - Brasil} (CAPES) -- Finance Code 001. EMT acknowledges support from UTFPR.}

\dataavailability{The original data can be shared upon reasonable request to the corresponding author.}

\acknowledgments{The authors acknowledge the National Laboratory for Scientific Computing (LNCC/MCTI, Brazil) for providing HPC resources of the SDumont supercomputer, which have contributed to the research results reported within this paper.}

\conflictsofinterest{The authors declare no conflicts of interest. The funders had no role in the design of the study; in the collection, analyses, or interpretation of data; in the writing of the manuscript; or in the decision to publish the results.}

%%%%%%%%%%%%%%%%%%%%%%%%%%%%%%%%%%%%%%%%%%
\appendixtitles{yes}
\appendixstart
\appendix

\section[\appendixname~\thesection]{Isolated galaxy}
\label{app:isolated}

\begin{figure}[H]
\includegraphics[scale=0.95]{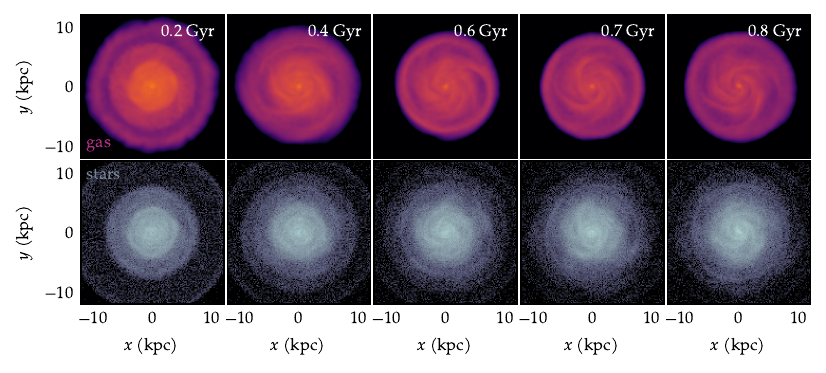}
\caption{The evolution of the isolated galaxy, showing the gas and stars at selected times.}
\label{015}
\end{figure}

\begin{figure}[H]
\includegraphics{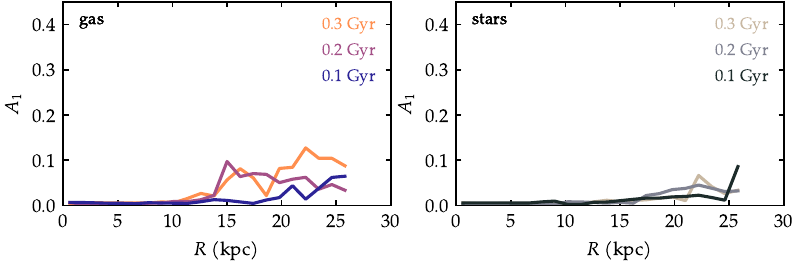}
\caption{Same as Fig.~\ref{fig4}, but for the isolated galaxy.}
\label{016}
\end{figure}

An isolated galaxy could in principle develop some degree of asymmetry due to its own internal secular evolution. To ensure that the peaks of $A_1$ seen in Fig.~\ref{fig4} are indeed caused by the ram pressure unwinding, the same measurements were performed using an isolated galaxy. This galaxy is an identical initial condition, but evolved in a vacuum, i.e.~in the absence of the ICM environment.
The morphological evolution of the isolated galaxy is seen in Fig.~\ref{015}. It undergoes some numerical transients in the first moments. The isolated galaxy does develop some mild intrinsic asymmetries detectable via the $A_1$ (Fig.~\ref{016}), but they are considerably smaller than in the stripped galaxy. This confirms that the peaks of $A_1$ in Fig.~\ref{fig4} are indeed capturing the asymmetries that were seen in the gas morphology.

\section[\appendixname~\thesection]{The Radon transform}
\label{app:radon}

The Radon transform of a function $f(x,y)$ in two dimensions is defined as the line integrals of $f(x,y)$ over all lines $L$ (see \cite{Krone2013, Stark2018} and references therein):
\begin{equation}
\tilde{f} = \int_Lf(x,y)ds,
\end{equation}
It takes a function defined in two-dimensional space (typically an image), and transforms it into $(p,\theta)$ space:
\begin{equation}
\label{eq:radon}
\tilde{f}(p,\phi) = \int_{-\infty}^{\infty}f(p\cos\phi - s\sin\phi, p\sin\phi + s\cos\phi)ds,
\end{equation}
corresponding to a set of projections of that function along different directions, with each value equal to the line integral of the function over that line.

\begin{figure}[H]
\includegraphics[scale=0.85]{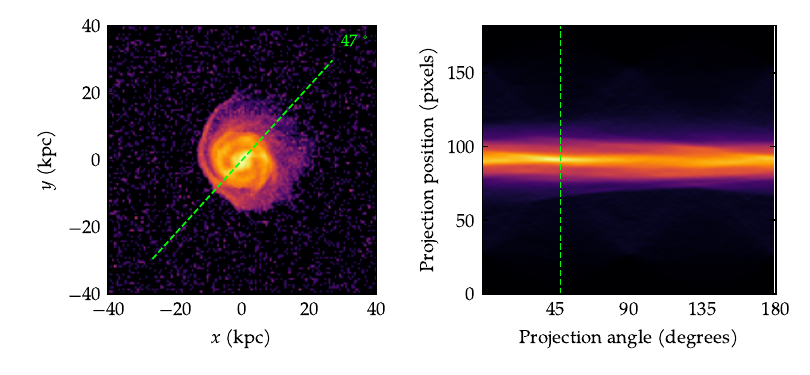}
\caption{Illustrative example of the application of the Radon transform. Left: Gas density map of the galaxy at 0.22\,Gyr. Right: Sinogram showing the result of the Radon transform as a function of projection angle. The dashed green lines correpond to the maximum response (right) and thus to the orientation of the galaxy (left) in this example.}
\label{010}
\end{figure}

When applied to the image of a galaxy, the angle for which the Radon transform yields the highest response provides a good estimation for the direction of greatest elongation. An illustrative example is shown in Fig.~\ref{010}.

%%%%%%%%%%%%%%%%%%%%%%%%%%%%%%%%%%%%%%%%%%
\begin{adjustwidth}{-\extralength}{0cm}

\reftitle{References}
\bibliography{paper}

\PublishersNote{}
\end{adjustwidth}

\end{document}